\documentclass[sigconf]{acmart}

\def\BibTeX{{\rm B\kern-.05em{\sc i\kern-.025em b}\kern-.08emT\kern-.1667em\lower.7ex\hbox{E}\kern-.125emX}}
    
\begin{document}

\copyrightyear{2019} 
\acmYear{2019} 
\acmConference[FDG '19]{The Fourteenth International Conference on the Foundations of Digital Games}{August 26--30, 2019}{San Luis Obispo, CA, USA}
\acmBooktitle{The Fourteenth International Conference on the Foundations of Digital Games (FDG '19), August 26--30, 2019, San Luis Obispo, CA, USA}
\acmPrice{15.00}
\acmDOI{10.1145/3337722.3341844}
\acmISBN{978-1-4503-7217-6/19/08}

%
\title{Introducing: The Game Jam License}

%

 \author{Gorm Lai}
\affiliation{%
 \institution{Goldsmiths, University of London}
 \city{New Cross, London}
 \country{United Kingdom}}
\email{glai001@gold.ac.uk}
 
 \author{Kai Erenli}
\affiliation{%
 \institution{University of Applied Sciences BFI Vienna}
 \streetaddress{Maria Jacobigasse 1}
 \city{Vienna}
 \country{Austria}}
 \email{kai.erenli@fh-vie.ac.at}
 
\author{Foaad Khosmood}
\affiliation{%
  \institution{Cal Poly}
  \streetaddress{1 Grand Av.}
  \city{San Luis Obispo}
  \state{California}
  \country{USA}}
  \email{foaad@calpoly.edu}
  
 \author{William Latham}
\affiliation{%
 \institution{Goldsmiths, University of London}
 \city{New Cross, London}
 \country{United Kingdom}}
\email{w.latham@gold.ac.uk}

%
\renewcommand{\shortauthors}{Lai, et al.}

%
\begin{abstract}
Since their inception at the Indie Game Jam\footnote{see: http://indiegamejam.com/igj0/index.html} in 2002, a significant part of game jams has been knowledge sharing and showcasing ideas and work to peers. While various licensing mechanisms have been used for game jams throughout the years, there has never been a licence uniquely designed for artifacts created during a game jam. In this paper, we present to the community the Game Jam License (GJL) which is designed to facilitate that sharing and knowledge transfer, while making sure the original creators retain commercial rights. The Global Game Jam\footnote{see: https://globalgamejam.org}, since 2009, strives to formalise sharing in a similar manner, by having jammers upload and license their creations under Creative Commons\footnote{see: https://creativecommons.org} Non Commercial Share Alike 3.0 free license. However, the CC family of licenses is not well suited for software. CC is not compatible with most other licenses, and introduces a legal grey area with the division between commercial and non-commercial use. Moreover, open source licences like GPL are well suited for source code, but not for art and design content. 
Instead the GJL presented in this paper, aims to uphold the original ideas of game jams (sharing and knowledge transfer), while still allowing the original team to hold on to all rights to their creation, without any of the deficiencies of the CC family of licenses.
\end{abstract}

%
%

%
\keywords{game jams, licenses, open source, creative commons, global game jam, copyright, sharing, law}

%
\maketitle

\section{Introduction}
\label{sec:introduction}
On-site game jams have been about sharing since their inception at the very first Indie Game Jam (2002), where game designers shared ideas and showed off their creations to each other while gathered in a central place. The Nordic Game Jam\footnote{see https://www.nordicgamejam.com} built further on this idea, and in the early years instructed jammers to upload and share source code and assets from their jam creation. The Global Game Jam has formalized this process by requiring jammers to upload a Creative Commons (CC) Non Commercial Share Alike 3.0 free license~\cite{2019:CCShareAlike30} along with their jam creations, consisting of executable, source code and assets. With this version of the CC licence, the original creators of the game retain all commercial rights, while those who download the game from globalgamejam.org can share and change the game, as long as they do not benefit commercially and remember to credit the original creators.\newline
While game jams are still a relatively new phenomenon, the rapid increase in size and scope of these events gives rise to new legal questions in need for more formalization. Over 47,000 participants created 9000 games during Global Game Jam 2019 alone\cite{2019:ggj2019}. Some interesting legal questions unique to game jams include the following: 
\begin{itemize}
    \item Can I share my game and still own it?
    \item Does participating in a game jam mean I can't commercialize my submissions?
    \item Can I bundle source and assets that I do not own into my game submission?
    \item Can I use closed-source commercial tools to build my game without compromising ownership of it?
    \item Am I free to download and modify previous game jam submissions?
    \item Can a single license cover source code, assets and art?
\end{itemize}
Existing practices such as using the CC license used by Global Game Jam have never been considered a good fit for source code. Moreover, open source licences are considered good fits for source code, but not for other content. We examined other choices, but no other license makes an effort to capture the interdisciplinary cooperative nature of game jams. Therefore, we investigate establishing a new licence that serves the global jamming community and addresses more of the relevant legal issues. Even this option has severe challenges: it must have fairly broad adoption to be useful.

\section{Previous Work}
\label{sec:previous}

As stated by \cite{2019:Bernd} certain products, such as video games, which have a strong software component but also integrate and create artistic works with an original element, in the meaning of the Directive 2001/29/EC (InfoSoc Directive)\cite{2001:EU} and as interpreted by the Court of Justice of the European Union (CJEU)\cite{2014:Leistner} could pose problems with the respect to the applicable law. Therefore, it was asked if video games should be considered solely under Directive 2009/24/EC (Software Directive)\cite{2009:EU}, due to the originality of the underlying source code or should they be considered as hybrid works under the InfoSoc Directive due to their original nature including artistic expressions and narratives. The CJEU stated in Nintendo vs PC Box \cite{2014:EU}, that video games fall within the scope of application of the InfoSoc Directive because they
\begin{quote}
    constitute complex matter comprising not only a computer program but also graphic and sound elements, which, although encrypted in computer language, have a unique creative value which cannot be reduced to that encryption. In so far as the parts of a video game, in this case, the graphic and sound elements, are part of its originality, they are protected, together with the entire work, by copyright in the context of the system established by Directive 2001/29.
\end{quote}

Keeping this in mind, it became apparent, that there is no license publicly available, which addresses the hybrid factor of video games in satisfactory way.

GJL is a license that is idealistic in nature, as it is essentially an attempt at reformulating the sharing spirit of the Global Game Jam into legal language. However, GJL is not the first attempt at doing such a thing. Richard Stallman famously created the Gnu General Public License (GPL) in 1989 to ensure that ``the users have the freedom to run, copy, distribute, study, change and improve the software''~\cite{Gay:2002:FSF}. The GPL license itself is predated by the GNU Manifesto~\cite{Stallman:1985:GNU}, which is a call for developers to join Stallman in creating free\footnote{Free is here meant in the famous quote ~\cite{Gay:2002:FSF} of ``To understand the concept, one should think of \textit{free} as in \textit{free speech}, not as in \textit{free beer}.''} UNIX compatible software. \newline
The spirit of the GPL license, where free is to be understood as ``freedom'' not ``price'', is also one of the primary motivators of the GJL, but we have a modern twist: Anyone can download, learn from and change the Standard version, but only the original creators are allowed to profit from the creation of the game and distribute subsequent closed versions. GJL still restricts the user's freedom in some ways. As Stallman argues, as he does in \cite{Stallman:2009:VWO}, that most open source licenses are not truly free, we believe his arguments would also apply to GJL.

The CC family of licenses~\cite{2019:CCLicenses} solve a similar problem, however as the CC FAQ~\cite{2019:CCFAQ} mentions, the license is not recommended to use for source code as it is incompatible with other licenses and does not deal with issues such as software patents. These patent apply to source code, but not to most other content e.g. images, sound, assets and models. Specifically, the CC FAQ page states:
\begin{quote}
    We recommend against using Creative Commons licenses for software. Instead, we strongly encourage you to use one of the very good software licenses which are already available ~\cite{2019:CCFAQ}.
\end{quote}

As is evident from the increased proliferation of online source code repositories such as Github\cite{2019:Github}, GitLab~\cite{2019:Gitlab} and BitBucket~\cite{2019:BitBucket}, the use of open source software is becoming ever more widespread. According to~\cite{2019:GithubMilestones}, Github reached 30 million developers in July 2018, and it hosts more than a staggering 2,800,000,000 lines of code by December 2018.\newline
In \cite{Orlando:2011:CL}, Guevara-Villalobos suggests that we are ``witnessing the configuration of communities of production as a means by which
developers seek to regain creative control over of their own labour''. Indie game developers form communities where they share ideas, learnings and resources. Game jams, such as the Global Game Jam, play an essential role as these creative spaces.

\subsection{Commercial vs Non-Commercial}
In order to provide jammers with some kind of legal framework, GGJ chose the CC, as it is a license that provides an explicit distinction between commercial and non-commercial use. Leaving aside the concerns about CC compatibility with software, this distinction seems useful for the actual artifact created during the jam. However CC is still inadequate in addressing one of the biggest concerns among game jammers: Are jammers allowed to monetize their projects after the jam is concluded? Unfortunately the answer to this question cannot be a simple ``yes'' or ``no'' as there are many aspects to be taken into (legal) consideration, e.g. is it considered ``commercial'' if a charity or non-profits sells the game jam package, or the package is used to get a grant or win a competition? Or what if a modified version is worked on for money or in-kind compensation? What if  the package is given away free as part of marketing campaign meant to promote a different product? The answer of ``it depends'' seems to be the most accurate one for this non-exhaustive list of example questions that demonstrates the legal distinction between commercial and non-commercial, but it is not a clear one.

\subsection{Patents}

The ability to patent software has caused a great discussion among software developers, lawyers and industry representatives. The legal framework differs globally. For example while the ability to patent software exists in the USA, it does not in Europe. Yet, many software licenses deal with patents to cover the issue. The CC is not among those licences as it does not address patents at all. Even though, Source code does seem unique in this context when compared to the rest of the content in the game jam package, and even though the package can be covered by copyright, the source code itself could be used for patent-able inventions and jammers should be aware of this fact. 

\subsection{The Incompatibility Issue}

It is evident, that except for the CC0 Public Domain Dedication license ~\cite{2019:CCPublicDomain}, and the CC Attribution-ShareAlike (BY-SA) version 4.0~\cite{2019:CCShareAlike40} which is one-way compatible with the GPL license~\cite{Gay:2002:FSF}, the CC licenses are not compatible with other open source licenses. Thus, the ability to link game jam content with content covered by such open source licences is limited and is considered as an unnecessary barrier by game jam organizers.

\section{Solutions}

Having discovered many problems and potential legal incompatibilities with existing practices, the authors resolve to come up with a solution. Three options are considered: splitting the package, using an existing license and developing a new license.

\subsection{Splitting the Package}
It can be argued that instead of creating a new licence, one solution is simply to split the game jam package into 2 packages: one package consisting solely of the source code and another package for the rest (mainly content). The latter will be licensed under the CC license, and the source code, under a more standard open source license that adheres to the main ideas of sharing while still crediting the original authors. Examples are the MIT~\cite{2019:MITLicense} or Artistic~\cite{2019:ArtisticLicense} license.
Instead of uploading one compressed file, jammers would simply upload two files, each falling under a different license.\newline

In theory, this idea works well, but in practice, the two packages always belong together, and one does not work without the other, so it becomes difficult to keep things conceptually separate.\newline

Additionally, not all game making tools allow the jammer to separate code and content to this degree. In fact, for most game developers it makes more sense to consider code as content. For example, is a blueprint made in Unreal Engine, code or content? How does one pull out such a blueprint and make it easily merge-able? What about a Unity prefab? Other tools, like Game Maker also make it difficult to separate the code.\newline
We conclude that while separating code and content into different packages might work for the traditional game developer using c++ or similar, for many modern game developers using game engines, separating the packages out can be impractical, confusing and burdensome.

\subsection{Using another License}
Instead of creating a new license, we first need to check if an already existing license can work as a substitute. We seek a license that:
\begin{itemize}
    \item covers patents
    \item is compatible with other major licenses
    \item does not require the creators of the package to sign away any rights to the software, commercial or otherwise
    \item makes sure sharing is permitted and encouraged, by allowing anyone to download the standard version and learn from its contents
\end{itemize}

The Creative Commons Share-Alike 3.0~\cite{2019:CCShareAlike30} has served the Global Game Jam well so far. However, as described in section~\ref{sec:introduction} and ~\ref{sec:previous}, the current license does not address the patent issue, and it is generally not recommended for software. 
The GPL license~\cite{Gay:2002:FSF} is great for sharing, but it does not allow for keeping a closed version of the code, unless the authors take great care in applying a license after they have forked the Standard Version, and makes it difficult to apply the later changes  to different forks of the Game Jam Package.
On the other hand, licenses such as the BSD~\cite{2019:BSDLicense} and MIT~\cite{2019:MITLicense} are too unrestricted, as they allow a user of the game jam package to make a modified version and sell it for profit, while not requiring that user to disclose and share the changes with the public.
In general, most licenses other than CC, can be divided into those that are Copyleft~\cite{2019:Copyleft} and those that are not. A Copyleft license such as the GPL, require later versions of licensed code to be licensed under similar term or stricter terms. The MIT and BSD licenses are not Copyleft.

\subsection{Developing a New License}
A new license is clearly required that fulfills the unique needs of the game jam movement. However, if this new license is to be successful and widely adopted, it should have a clear governing authority. Such an authority is not strictly necessary for licenses as the text of the license should be self explanatory, but in practice, new challenges, technological innovations and legal landscapes could necessitate changes to future versions of this license. Therefore, we strongly suggest an independent governing authority, made up of interested game jam community leaders, be convened to oversee the development and distribution of this license, and be the custodian of its documentation. 
Our license version 0.1 is reproduced in Appendix~\ref{Appendix:TheLicense}. 


\section{Conclusion And Future Work}
In this paper we examine the issues of content and software licensing as related to game jam activities. The license presented here, is meant to be a conversation starter. New licenses are rare and require years of promotion for wide adoption. In general the consensus among the authors is that we must not create a brand new license (``yet another license'') unless uniquely and specifically called for.\newline

The authors believe the case can be made for a new license. We invite the community to contact us for feedback and suggestions. Our proposal and first version is available in Appendix~\ref{Appendix:TheLicense}. 

%
\begin{acks}
This work was in part funded by the EPSRC Centre for Doctoral Training in Intelligent Games and Game Intelligence (IGGI) EP/L015846/1.
\end{acks}

%
\bibliographystyle{ACM-Reference-Format}
\bibliography{gamejamlicense}

%
\appendix
\onecolumn
\section{The License version 0.1}
\label{Appendix:TheLicense}

\textit{Everyone is permitted to copy and distribute verbatim copies of this license document, without modifications.}

\subsection{Preamble}

This license establishes the terms under which a given Game Jam Package may be distributed, copied, modified, developed and/or redistributed. The intent is that every Participant of a Game Jam Event can contribute to the Jam under a legal framework while still keeping the Game Jam Package available and alive.\newline

You are always permitted to decide wholly outside of this license directly with the Copyright Holder of a given Game Jam Package. Nevertheless, the Game Jam Package developed at any participating Game Jam Event is covered by this license and you agree to comply with the rules of the license.

\subsection{Definitions}
\begin{itemize}
    \item ``Copyright Holder'' means the individual(s) Contributors named in the copyright notice for the entire Game Jam Package.
    \item ``Contributor'' means any party that has contributed code or other material to the Package, in accordance with the Copyright Holder's procedures.
    \item ``You'' and ``your'' means any person who would like to copy, distribute, or modify the Package
    \item ``Intellectual Property'' means the intellectual property rights set out in the schedule to this agreement.
    \item ``Game Jam Package'' means the collection of (not limited to) source and binary files, symbols, designs, assets, art work, logos, characters and sounds developed at and during any Game Jam Event.
    \item ``Distribute'' means providing a copy of the Package or making it accessible to anyone else.
    \item ``Standard Version'' refers to the unmodified Game Jam Package.
    \item ``Modified Version'' means the modified Game Jam Package whether modified by the Copyright Holders or by a third party or person.
    \item ``Organizer'' means any member of the organization team of any Game Jam.
    \item ``Original License'' means this Game Jam License as distributed with the Standard Version of the Package, in its current version or as Game Jam Organizers may modify it in the future.
    \item ``Source'' form means the source code, documentation source, and configuration files for the Game Jam Package.
    \item ``Compiled'' form means the compiled bytecode, object code, binary, or any other form resulting from mechanical transformation or translation of the Source form.
\end{itemize}
\subsection{Licence Grants}

\subsubsection{Rights \& Game Ownership, Sequels, Expansions \& Ports}

Copyright Holders of each Game Jam Package grant each other a royalty-free, charge-free, perpetual, irrevocable, non-exclusive, worldwide, transferable license to use, modify, share, make, run, make-available, redistribute and propagate the Game Package without Copyright Holder's prior written permission as long as the following items are included:
\begin{itemize}
    \item a copy of the license
    \item attribution to the original creators (the Copyright Notice), unless specifically waived by creators
    \item name and date of the originating Game Jam event (as specified in the Copyright Notice)
\end{itemize}

You are permitted to use the Standard Version and create and use Modified Versions for any purpose without restriction, provided that you do not Distribute the Modified Version.

\subsubsection{Source Code, Development Tools, Third Party Software \& Engine Ownership}

Contributor will comply with all applicable law, regulation, and third party rights (including without limitation laws regarding the import or export of data or software, privacy, and local laws). Contributor will not encourage or promote illegal activity or violation of third party rights.\newline

Contributor will not incorporate any pre-existing invention, improvement,
development, concept, discovery, copyright protect work, or other proprietary information not owned (\textit{Contributor would need the ability to fully license these materials to be compliant with this license}) by Contributor

\subsubsection{Indemnification}

Contributor agrees to indemnify and hold harmless the Contributors and Copyright Holders of the Game Jam Package from and against all taxes, losses, damages, liabilities, costs and expenses, including attorneys' fees and other legal expenses, arising directly or indirectly from or in connection with
\begin{itemize}
    \item any reckless or intentionally wrongful act of Contributor or Contributor's
    \item any breach by the Contributor of any of the covenants and warranties contained in this Agreement
    \item any violation or claimed violation of a third party's rights resulting in whole or in part.
\end{itemize}

\subsubsection{Limitation of Remedies}

The Contributor expressly agrees that throughout the duration of this Agreement, in the event of any default of any terms of this Agreement, Contributor's only remedy will be an action at law for damages and in no event shall either the Contributor be entitled to rescind this Agreement or to receive any injunctive or other equitable relief or to restrain the distribution, exhibition, advertising or other exploitation of the
Game Jam Package

\subsubsection{Confidentiality \& Privacy Issues (revise for appropriate frame)}

No Contributor may use, disclose or make available to any third party the other Party's Confidential Information, unless such use or disclosure is done in accordance with the terms of this Agreement.\newline

The Contributor hereby consents and acknowledges that different countries where a Game Jam takes place may not have laws in place to adequately protect his data and his privacy.

\subsubsection{Liability \& Warranties}

The Contributor acknowledges and agrees that neither the Organizers nor its board members will be liable for any loss or damage arising out of or resulting from Contributor's provision of the Game Jam Package under this Agreement, or any use of the Game Jam Package by the Contributor; and Contributor hereby releases the Organizers to the fullest extent from any such liability, loss, damage or claim.\newline

THE GAME JAM PACKAGE IS PROVIDED BY THE COPYRIGHT HOLDER AND CONTRIBUTORS ``AS IS'' AND WITHOUT ANY EXPRESS OR IMPLIED WARRANTIES. THE IMPLIED WARRANTIES OF MERCHANTABILITY, FITNESS FOR A PARTICULAR PURPOSE, OR NON-INFRINGEMENT ARE DISCLAIMED TO THE EXTENT PERMITTED BY YOUR LOCAL LAW. UNLESS REQUIRED BY LAW, NO COPYRIGHT HOLDER OR CONTRIBUTOR WILL BE LIABLE FOR ANY DIRECT, INDIRECT, INCIDENTAL, OR CONSEQUENTIAL DAMAGES ARISING IN ANY WAY OUT OF THE USE OF THE GAME JAM PACKAGE, EVEN IF ADVISED OF THE POSSIBILITY OF SUCH DAMAGE.

\subsubsection{Term \& Termination}

You may not propagate or modify a Game Jam Package except as expressly provided under this License. Any attempt otherwise to propagate or modify it is void, and will automatically terminate your rights under this License.

\subsubsection{General Provisions}

Any use, modification, and distribution of the Standard or Modified Versions is governed by this Game Jam License. By using, modifying or distributing the Package, you accept this license. Do not use, modify, or distribute the Game Jam Package, if you do not accept this license.\newline

If your Modified Version has been derived from a Modified Version made by someone other than you, you are nevertheless required to ensure that your Modified Version complies with the requirements of this license.\newline

This license does not grant you the right to use any trademark, service mark, tradename, or logo of any Copyright Holder.\newline

EACH CONTRIBUTOR ACKNOWLEDGES THAT, IN EXECUTING THIS AGREEMENT, CONTRIBUTOR HAS HAD THE OPPORTUNITY TO SEEK THE ADVICE OF INDEPENDENT LEGAL COUNSEL, AND HAS READ AND UNDERSTOOD ALL OF THE TERMS AND PROVISIONS OF THIS AGREEMENT.
THIS AGREEMENT SHALL NOT BE CONSTRUED AGAINST ANY PARTY BY REASON OF THE DRAFTING OR PREPARATION HEREOF.\newline

This license includes the non-exclusive, worldwide, free-of-charge patent license to make, have made, use, offer to sell, sell, import and otherwise transfer the Game Jam Package with respect to any patent claims licensable by the Copyright Holder that are necessarily infringed by the Package. If you institute patent litigation (including a cross-claim or counterclaim) against any party alleging that the Game Jam Package constitutes direct or contributory patent infringement, then this Game Jam License to you shall terminate on the date that such litigation is filed.\newline

This Agreement will be construed by and governed in accordance with the laws of [COUNTRY]. The Parties submit to exclusive jurisdiction of the courts of
[COUNTRY].\newline

If any provision of this Agreement is found to be illegal or unenforceable, the other provisions shall remain effective and enforceable to the greatest extent permitted by law.

\end{document}